# Ambipolar charge transport in quasi-free-standing monolayer graphene on SiC obtained by gold intercalation

*Kyung Ho Kim,[1] Hans He,[1] Claudia Struzzi,[2] Alexei Zakharov,[2] Cristina Giusca,[3] Alexander Tzalenchuk,[3,4] Rositsa Yakimova,[5] Sergey Kubatkin,[1] Samuel Lara-Avila[1,3,\*]*

[1]Department of Microtechnology and Nanoscience, Chalmers University of Technology, SE-412 96, Gothenburg, Sweden

[2]MAX IV Laboratory, 221 00, Lund, Sweden

[3]National Physical Laboratory, Hampton Road, Teddington TW11 0LW, UK

[4]Royal Holloway, University of London, Egham TW20 0EX, UK.

[5]Department of Physics, Chemistry and Biology, Linkoping University, 581 83 Linköping, Sweden.

**We present a study of quasi-free-standing monolayer graphene obtained by intercalation of Au atoms at the interface between the carbon buffer layer (Bu-L) and the silicon-terminated face (0001) of 4H-silicon carbide. Au intercalation is achieved by deposition of atomically thin Au on the Bu-L followed by annealing at 850 °C in an Argon atmosphere. We explore the intercalation of Au and decoupling of the Bu-L into quasi-free-standing monolayer graphene by surface science characterizations and electron transport in top-gated electronic devices. By gate-dependent magnetotransport we find that the Au-intercalated buffer layer displays all properties of monolayer graphene, namely gate tunable ambipolar transport across the Dirac point, and n- or p-type doping depending on the Au content.**

Epitaxial graphene grown on SiC (epigraphene) is a promising route for scalable graphene electronics. The silicon-terminated face (0001) offers the possibility to grow continuous, single-crystal graphene over wafer scale, as demonstrated for hexagonal SiC polytypes 4H and 6H.[1,2] A challenge for epigraphene electronics is that the monolayer on the silicon face is heavily n-type doped, with a typical intrinsic doping n ~ $1 \times 10^{13}$ electrons/cm$^{-2}$ pinned by the SiC substrate,[3] and this complicates tuning the carrier density of epigraphene with e.g. electrostatic gates.[3–5] The high doping originates from the structure of the interface between SiC and the epigraphene, where an interface layer, the so called carbon buffer layer (Bu-L), serves as a source of donor-like states that result in n-type doping.[6]

A route aiming at controlling the graphene-SiC interaction is by intercalation of hydrogen at the epigraphene-SiC interface.[7,8] In this intercalation process, epigraphene is subjected to a thermal annealing step in a hydrogen-containing atmosphere. At high temperatures, hydrogen atoms migrate into the interface between the Bu-L and SiC, breaking the Si-C bonds and saturating the Si dangling bonds resulting in the lifting of the graphene/Bu-L into quasi-free-standing (qFS) bilayer graphene.[7,8] The same process can be applied to the Bu-L which, remarkably, recovers all attributes of monolayer graphene (MLG) after hydrogen intercalation.[7–12] Yet, the resulting quasi-free-standing monolayer might still suffer from strong interaction with SiC, due to spontaneous polarization of the hexagonal substrate, leading to heavy p-type doping.[6] Besides hydrogen, other atomic species can be intercalated.[13] Among all intercalants, Au has shown the possibility to result in n-type or p-type doped epigraphene depending on the amount of gold that is intercalated.[14] In addition to controlling the doping type, Au is attractive because it could lead to enhancing spin-orbit interaction (SOI) in graphene, which holds promises for a variety of spin-related phenomena in graphene.[15–18] From a practical point of view, the chemical inertness of Au is also attractive, because it can enable the ex-situ processing of intercalated epigraphene into devices.

We report electron transport studies complemented by surface science analysis of quasi-free-standing graphene obtained by Au intercalation at the Bu-L/SiC interface (Fig. 1a). We use the epitaxial Bu-L grown on 4H-SiC (0001) substrates and decoupled it from the SiC substrate by deposition of Au atoms on the surface followed by a thermal drive-in at T = 850 °C. The strong interaction between Au and the Bu-L allows for smooth deposition of Au monolayers, without Au agglomeration into clusters.[19] After Au intercalation, we study electronic transport properties of top gated devices and find that Au-intercalated Bu-L displays properties of monolayer graphene, with a zero-gate doping level located at n = 0.7 – 1.2 × $10^{12}$ cm$^{-2}$.

**Results and Discussion**

To prepare Au-intercalated quasi-free-standing graphene, we start with 7 mm × 7 mm SiC/Bu-L substrates onto which we deposit Au on about half of the chip area (5 mm × 2.5 mm) (Fig. 1b). We have studied deposition of two different Au thickness, $t_{Au}$ = 4 Å and 8 Å, in two SiC/Bu-L substrates. Our deposition conditions result in a cluster-free Au layer on the surface, as evidenced by atomic force microscopy (See Fig. S1),[19] with an RMS roughness of 1 Å. Despite the homogeneous coverage of Au on the Bu-L surface, such layers are electrically not conductive.[19] After Au deposition, the intercalation step at T = 850 °C results in an electrically conductive surface, showing resistance values of the order of few kOhms (~3 – 20 kΩ). This onset of electrical conductivity occurs not only on the Au deposited area, but on the rest of the surface as well, where Au was originally not present, pointing to diffusion of Au across the surface. Together with the change in electrical conductance, optical inspection[20] of the chip allows to see that the thermal intercalation step leads to a modification in the transparency of the surface, which becomes more opaque millimeters away from the Au deposited area (Fig 1b). We inspected the surface with Raman spectroscopy with a laser wavelength of 638 nm, which readily shows the emergence of the graphene 2D peak at 2662 cm$^{-1}$ and G peak at 1583 cm$^{-1}$ everywhere on the surface, with a FWHM = 63 cm$^{-1}$ for the dark and electrically conductive area, and FWHM = 68 cm$^{-1}$ on the Au-deposited area (Fig. 1(c,d)).[14] Both the Raman spectra and the electrical conductivity of the surface after the annealing step serve as a strong indication that the Bu-L has decoupled from the substrate and transformed into monolayer graphene everywhere on the surface that becomes more opaque. Scanning tunneling microscopy (STM) analysis shows that the Au-deposited area displays a granular morphology consistent with the presence of Au clusters on the surface (Fig. 1e), which gradually disappear as one moves away from the border with the diffused Au area. In the Au-diffused area, the surface appears clean and free of Au clusters, revealing a terraced surface similar to the typical SiC/Bu-L morphology (Fig.1f). As we describe below, the dark and electrically conductive area correspond to areas where Au has diffused over the surface and intercalated at the buffer-SiC interface, and we name this area the Au-diffused region as shown in Fig. 1b.

Studies at synchrotron facilities confirm the nature of Bu-L on the Au-deposited and Au-diffused area. Fig. 2a is the low energy electron microcopy (LEEM) image at the boundary of the Au-deposited and Au-diffused areas after thermal drive-in of 4 Å Au. The contrast in the two regions is due to the slightly different amount of intercalated gold (the darker grey color scale corresponds to lower gold content). Decoupling of the Bu-L occurs on both the gold-deposited and gold-diffused areas, which is confirmed through the quenching of the $6\sqrt{3} \times 6\sqrt{3}$ R30° pattern in low energy electron diffraction (LEED) (Fig. 2(b,c), see Fig. S4). The diffractograms also show the hexagonal honeycomb structure of graphene after the thermal drive-in, visible for both gold-deposited and gold-diffused areas. The LEEM intensity IV curves from both sides of the boundary are shown in Fig. 2d: the dip at 5.5 eV is resulting from the intercalation and consequent formation of free-standing graphene. The intercalation of Au as the reason for the decoupling is further confirmed by the x-ray photoelectron microspectroscopy results (micro-XPS), showing the presence of Au in both the Au-deposited and Au-diffused areas and the gradual disappearance of Au signal far from the boundary (Fig. 2e). Another evidence of Au-intercalation on both sides of the boundary can be seen in the C1s micro-XPS spectra (Fig. 2f), in which a charge transfer from gold to graphene increases the separation between carbon peaks from graphene and the SiC substrate.[13]

To assess the electrical properties of the Au-intercalated quasi-free-standing graphene we have fabricated micro-sized devices on the two substrates ($t_{Au}$ = 4 Å, 8 Å). In total, we have studied 8 devices, placed on the Au-deposited and Au-diffused areas of the substrates. Devices were made by conventional electron beam lithography (EBL) and oxygen plasma etching. For gated devices, we have used dry-transferred hexagonal boron nitride (h-BN) (thickness ~20 nm) followed by atomic layer deposition of $Al_2O_3$ (38 nm) as a dielectric, and Ti/Au as a gate electrode. Fig. 3(a–c) depicts the schematic structure of the top-gated devices in the top and edge contact configurations together with an optical micrograph of D1. To avoid further processing steps, h-BN was not patterned, and the geometry of the devices is dictated by the shape of the transferred h-BN flakes. Magneto transport properties of the devices were measured by the van der Pauw method in a gas flow cryostat down to T = 2 K. We quantified the Hall carrier density and mobility as $n_H$ = $1/eR_H$ and $\mu_H$ = $R_H \times \sigma_{xx}$, respectively.

We found that devices made on the Au-deposited area are insensitive to the gate voltage, likely due to screening of the electric field by the Au layer present directly atop the graphene layer. Gate response is only observed in those devices fabricated on the Au-diffused area, where Au is absent on top of the graphene layer according to STM scans. Fig. 3d displays the top gate dependence on electron transport of devices D1 and D2 fabricated on the Au-diffused area, at T = 2 K, with top and edge contact, respectively, showing ambipolar transport across what we attribute to be a Dirac point (DP), at gate voltages $V_g$ = $V_D$ = -1.4 V (-5.2 V) for D1 (D2). From the gate response, both devices are mildly n-doped. The gate allows us to tune the carrier density at the level $\alpha$ = $dn_H/dV_g$ ~ 1.8 × $10^{11}$ $cm^{-2}$/V (D1) and 1.7 × $10^{11}$ $cm^{-2}$/V (D2), which agrees well with the dielectric thickness (~20 nm h-BN) and the dielectric constant of the dielectric materials, $\varepsilon_{h-BN}$ = 3.76 and $\varepsilon_{Al2O3}$ = 9.[21,22] Fig. 3e shows Hall carrier density of D1(D2) at T = 2

K at different gate voltages, where the type of the majority carrier shifts from electrons to holes at a gate voltage $V_g = V_D$. The Hall mobility for device D1 is $\mu_H = 200 - 250$ cm$^2$/Vs for holes and $\mu_H = 70 - 200$ cm$^2$/Vs for electrons; for device D2, somewhat larger mobilities are measured on the electron side, $\mu_H = 500 - 600$ cm$^2$/Vs. The gated devices allow us to compare the Hall carrier mobilities with those extracted by capacitive method using the equation $R_{xx} = \frac{1}{e\mu_C}(\frac{L}{W})\frac{1}{\sqrt{n_g^2+n_0^2}}$, where the e, $\mu_c$, L (W), $n_g = \alpha(V_g-V_D)$, $n_0$ is electron charge, field-effect mobility, length (width) of the sample, carrier density induced by gate, and residual carrier density, respectively.[23] This capacitor model gives $\mu_c = 270$ (560) cm$^2$/Vs and $n_0 = 6 \times 10^{11}$ ($7 \times 10^{11}$) electrons/cm$^2$ for D1 (D2), showing consistency between $\mu_H$ and $\mu_c$. Given the residual carrier density $n_0$ and the $V_D = -1.4$ V (-5.2 V) for D1 (D2), we estimate carrier density at zero gate voltage $n = \sqrt{n_g^2 + n_0^2} = 7 \times 10^{11}$ ($1.2 \times 10^{12}$ cm$^{-2}$) electrons/cm$^2$.

Table I shows a summary of transport properties measured in all the devices. In general, the zero-gate doping of Au-intercalated devices is p-type, $p = 5 \times 10^{12} - 2 \times 10^{13}$, being consistently higher in magnitude for the Au-deposited regions. The notable exception is for gated devices fabricated on the Au-diffused areas, which show n-type doping. According to Gierz et al.,[14] Au contents corresponding to 3/8 monolayer (Au-ML) and 1 Au-ML results in highly n-doped (n = 5 × 10$^{13}$ cm$^{-2}$) and slightly p-doped[15] (p = 7 × 10$^{11}$ cm$^{-2}$) MLG, respectively.[24] Therefore, the n-type (p-type) doping in Au-diffused (deposited) areas could arise from the Au contents less (higher) than 1 ML. However, we cannot exclude the possibility of doping of the decoupled MLG from environment during the fabrication processes. While gated devices allow us to explore the carrier mobility as a function of the gate voltage, the mobilities of other devices at a fixed carrier density do not exceed 100 cm$^2$/Vs. We note that the reported mobility of hydrogen intercalated Bu-L is of the order of 1000 cm$^2$/Vs,[9–12] suggesting the possibility that the relatively low mobility is intrinsic in Bu-L as a consequence of defects on the as-grown Bu-L[25] or of an imperfect intercalation process. Finally, measurements in magnetic field allow us to analyze the nature of microscopic scattering processes in Au-intercalated Bu-L. Fig. 3f shows the symmetric part of the longitudinal magneto conductivity ($\Delta\sigma_{xx} = \sigma_{xx}(B) - \sigma_{xx}(0)$) of D1 at T = 2 K (See Fig. S6). This measurement serves as a representative example of the low field magneto conductivity for all measured devices. To mention, the observation of positive magneto conductance in all devices rules out the enhancement of spin-orbit scattering in graphene by Au atoms present beneath and on top of graphene. With a gate-dependent momentum relaxation length $L_p = h/(2e^2\rho_{xx}\sqrt{n\pi})$ in the range 2 – 5 nm (3 – 8 nm) for D1 (D2), the curvature $\kappa$ of the low field magnetoconductance,[26,27] $\kappa = \frac{\partial^2 \sigma_{xx}}{\partial B^2}\Big|_{B=0} = \frac{16\pi}{3}\frac{e^2}{h}\left(\frac{D\tau_\varphi}{h/e}\right)^2$, allows us to quantify a phase-relaxation time in the range $\tau_\varphi = 0.7 - 3.4$ ps ($L_\varphi = 33 - 42$ nm) for D1 and $\tau_\varphi = 0.4 - 0.7$ ps ($L_\varphi = 30 - 39$ nm) for D2, being higher near the DP (see Fig. S7). With the short phase-relaxation time and the overall low mobility in the samples, we speculate that the absence of spin-orbit scattering might be related to disorder in Au-intercalated samples. In fact, it is only for devices fabricated on the substrate containing $t_{Au} = 4$ Å that the temperature dependence of conductivity is close to logarithmic (See Fig. S8), implying that the temperature dependence of these devices could be attributed to quantum corrections to the conductivity. However, if the phase-relaxation time is shorter than the spin-orbit scattering time $\tau_{SO} > \tau_\varphi$, the effects of spin-orbit scattering might not be observable in our samples under the measured conditions (down to T = 2 K). In general, the SOI enhancement of MLG in proximity to Au is affected by the structure of graphene-Au interfaces including hybridization of orbitals, graphene-Au distances, and position of Au respect to the graphene lattice.[15,28–31] Band calculations predict that the optimal configuration for SOI enhancement is that of Au atoms located at hollow sites of graphene.[15,28] Therefore, it is possible that the absence of spin-orbit scattering effects in our samples might be related to the Au atoms occupying random sites (i.e. disorder) at the Bu-L/SiC interface. Moreover, it appears that the excess of gold contributes to enhancing the disorder in the samples. The temperature dependence of devices fabricated with quasi-free-standing graphene with higher Au content ($t_{Au}$ = 8 Å) follows the characteristic dependence of granular metals[32] and variable range hopping[33]: $R(T) = R_0 \exp[(T_0/T)^{1/2}]$, with $T_0$ = 30 K – 100 K and $R_0$ = 3 kΩ – 9 kΩ for device D5, D6, and D7, and $R(T) = R_0 \exp[(T_0/T)^{1/4}]$, with $T_0$ = 320 K and $R_0$ = 7 kΩ for device D8, suggesting that excess gold introduces additional sources of momentum and energy relaxation of carriers.

**Conclusions**

In conclusion, we show that Au-intercalated quasi-free-standing monolayer graphene can be obtained by deposition of one or few gold monolayers on the Bu-L surface followed by annealing in Ar atmosphere. Au can diffuse under the Bu-L surface and this leads to diluted contents of Au millimeters away from the Au-deposited area. Au-intercalated quasi free standing monolayer graphene samples are stable in ambient conditions and allow for the ex-situ fabrication of devices and material characterization. By decoupling it from the substrate, the Au-intercalated monolayer graphene shows pronounced gate modulation of conductance, with a Dirac point at $V_D$ = -1.4 V and -5.2 V (carrier density at $V_g$ = 0 is n = 7 × 10$^{11}$ and 1.2 × 10$^{12}$ electrons/cm$^2$). Absence of SOI enhancement could be attributed to disorder in the intercalated Au layer. Future efforts in optimizing the thermal drive-in step will reveal if Au can be orderly assembled at the Bu-L/SiC interface, to enhance spin-orbit scattering in quasi-free-standing epigraphene on insulating SiC substrate.

**Experimental Methods**

**Growth of Bu-L on SiC:** The carbon buffer is an integral part of the epitaxial graphene-SiC material system and is the first to form when SiC substrate is exposed to high temperature (T > 1500 °C). More specifically, this is the carbon rich surface reconstruction (6√3 × 6√3) characteristic of Si face SiC at elevated temperatures. To prevent the growth of graphene and grow only the carbon zero-layer, here we used 7 × 7 mm$^2$ 4H-SiC substrates and applied gradual (inductive) heating in Argon atmosphere until T ≈ 1700 °C was reached and that kept for 30 seconds. Then the furnace was switched off and the samples were taken out at room temperature. Prior to growth, the chamber was pumped down to a base pressure of $P_0 = 1 \times 10^{-6}$ mbar in order to minimize oxygen contamination which is detrimental for a complete carbonization.

**Au deposition:** Metals were deposited by e-beam evaporation at base pressure $P_0 = 5 \times 10^{-7}$ mbar in a Lesker PVD 225 fitted with a custom-built substrate heater. Deposition on part of the substrate took place through a shadow mask, to avoid surface contamination. Before metal deposition, the substrate temperature is raised to 200 °C and kept constant for 5 minutes. The deposition rate is set to r = 1 Å/s, yielding a deposition time of 4 s and 8 s, for $t_{Au}$ = 4 Å and t = 8 Å, respectively.

**Intercalation:** Substrates were heated up in an atmosphere of ultra-pure Argon of $P_0$ = 800 mbar at T = 850 °C for 90 minutes. Lower temperatures resulted in partial intercalation of Au, while at higher temperatures the intercalated Au layer can escape the interface.

**LEEM and micro-XPS:** Low energy electron microscopy (LEEM) experiments were performed at MAXPEEM beamline at MAX IV synchrotron facility (1.5 GeV ring), Lund, Sweden. The beamline is housing aberration corrected spectroscopic photoemission and low energy electron microscope (SPELEEM). The microscope has a plethora of imaging modes, which is capable of delivering information on structural, chemical, electronic, and magnetic contrast at spatial resolutions in the nanometer range in one instrument. In addition to imaging modes, microspectroscopy measurements are possible allowing acquisition of micro-XPS spectra from an area of interest in the 1 to 10 μm range.

**STM:** The STM measurements were carried out at NPL on an ultra-high vacuum LT Nanoprobe Scienta Omicron system, base pressure $4 \times 10^{-11}$ mbar, at room temperature, with no prior surface conditioning of the sample. The images were recorded at a tunnelling current of 1 nA and a bias voltage of -0.4 V, using electrochemically etched W wires.

**Top gate fabrication:** To fabricate top gate devices, h-BN flakes dehydrated by baking at 200 °C for 20 min in nitrogen atmosphere were mechanically exfoliated on top of the Au intercalated Bu-L prebaked at 200 °C for 10 min in nitrogen. The exfoliated substrate was directly transferred to the furnace and annealed at 750 °C for 1 hour in vacuum to remove hydrocarbon impurities and water at the interface. D2 was edge-contacted by oxygen plasma etching followed by contacting with Ti/Au (5/80 nm) after EBL, and other devices were top-contacted by contacting with Ti/Au (5/80 nm) after EBL followed by oxygen plasma etching.

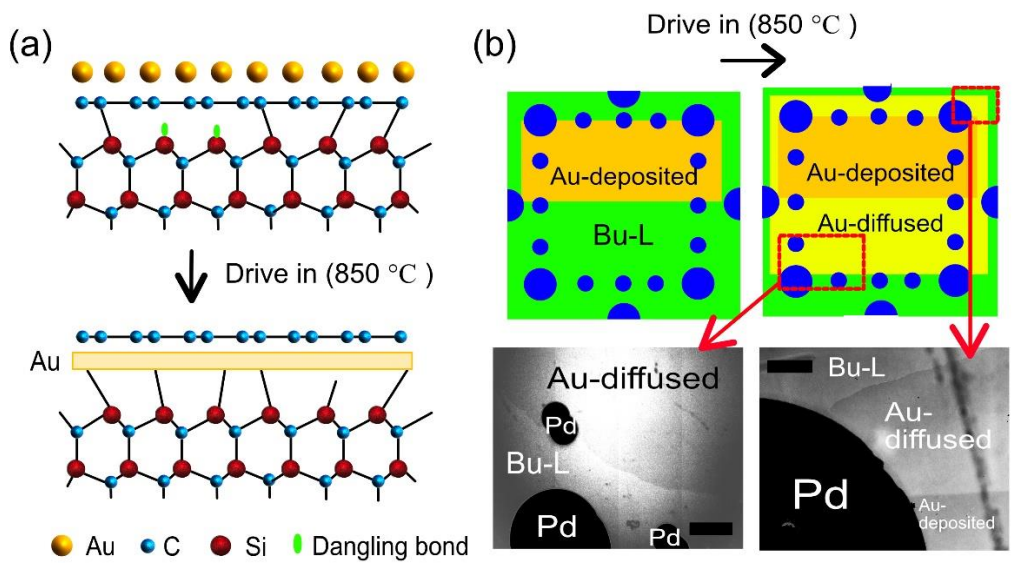

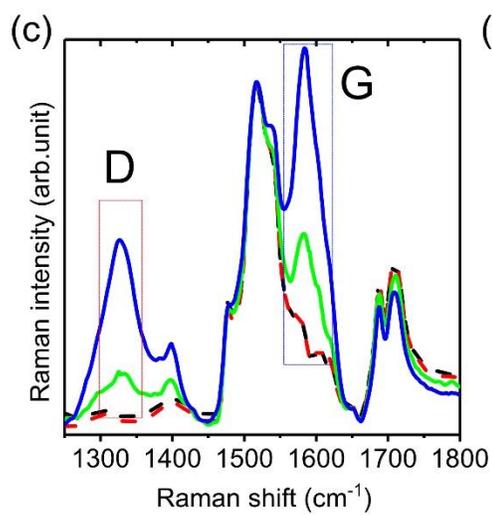
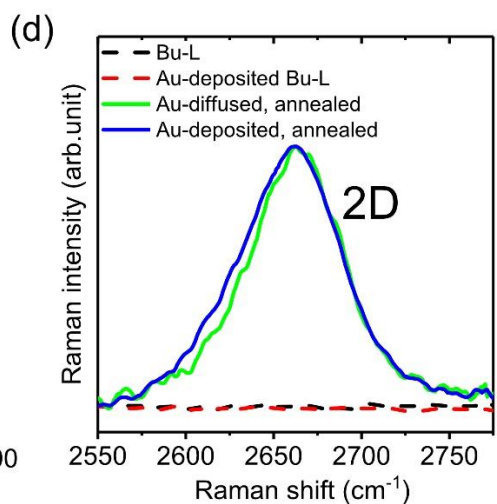

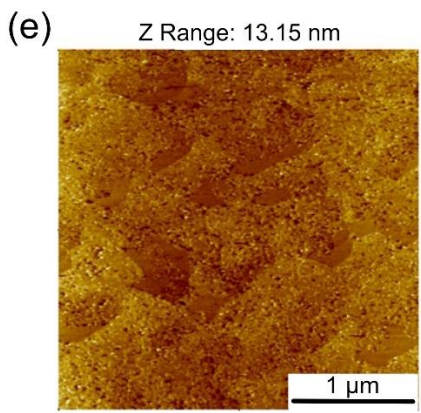
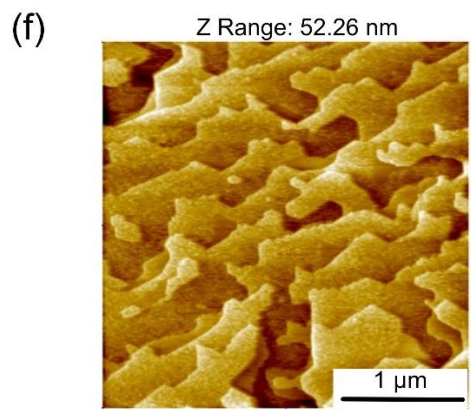

**Figure 1. Au-intercalated quasi free standing monolayer graphene** (a) A schematic of Au atoms on top of the Bu-L and intercalation of Au atoms after thermal drive, lifting the Bu-L from SiC (0001). (b) Schematics of deposition of Au atoms (yello) on the half the Bu-L substrate and after thermal drive, showing diffusion of Au (light yello, Au-diffused). The blue circles are Pd contacts. The bottom figures are optical micrographs of the areas in the red boxes, showing the diffusion of Au over mm lengths. The black circles are Pd contacts and the scale bar is 100 µm. (c,d) Raman D and G peaks (c) and 2D Peaks (d) of Bu-L (black), Au deposited Bu-L (red), Au-diffused area (green), and Au-deposited area after intercalation (blue), showing the emergence of graphene D, G, and 2D peaks after decoupling the Bu-L. (e,f) STM topography images after 4 Å Au intercalation, where the Au-deposited area shows Au clusters (e) and the Au-diffused area (f) is cluster-free with clear terraces on the surface.

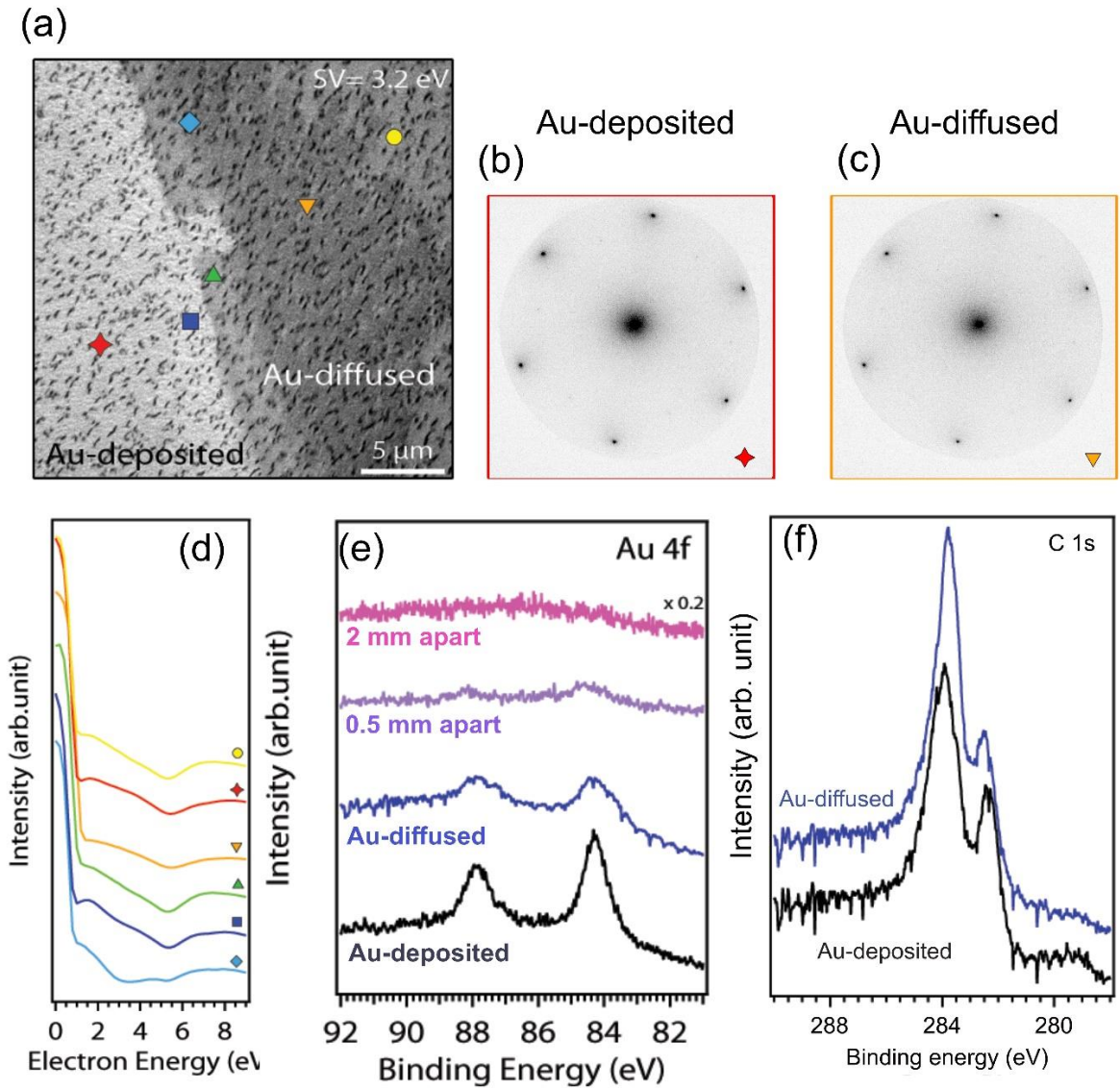

**Figure 2. Synchrotron studies of Au-intercalated quasi-free-standing monolayer graphene**

(a) Low energy electron microscopy (LEEM) image (Field of View = 25 μm, electron energy 3.2 eV) at the boundary of Au-deposited and Au- diffused areas (see Fig. 1b) after 4 Å Au intercalation. The black patches in the image were monolayer graphene (but now bilayer after intercalation). (b,c) Micro-low energy electron diffraction (LEED) patterns (SV = 45 eV, sampling area 1.5 µm) recorded from the marked areas shown in (a) with corresponding colors and shapes. Au-deposited (b) and diffused (c) that show only diffraction from monolayer graphene and no traces of 6√3×6√3R30° structure which is the fingerprint of the buffer layer.

(d) LEEM IV curves from both Au-deposited and Au-diffused areas recorded from the marked areas shown in (a) with corresponding colors and shapes. A dip in the IV LEEM at 5.5 eV is a result of intercalation and consequent formation of free-standing graphene. (e) Micro-XPS spectra (hv = 150 eV, sampling area 5 µm) of Au 4f at different areas show the presence of Au on Au-deposited (black) and diffused (blue) areas and a gradual disappearance of Au signal as the distance from the boundary increases; violet (0.5 mm), pink (2 mm) away respectively. (f) micro-XPS spectra (hv = 350 eV, sampling area 5 µm) of C 1s on Au-deposited (black) and diffused (blue) areas.

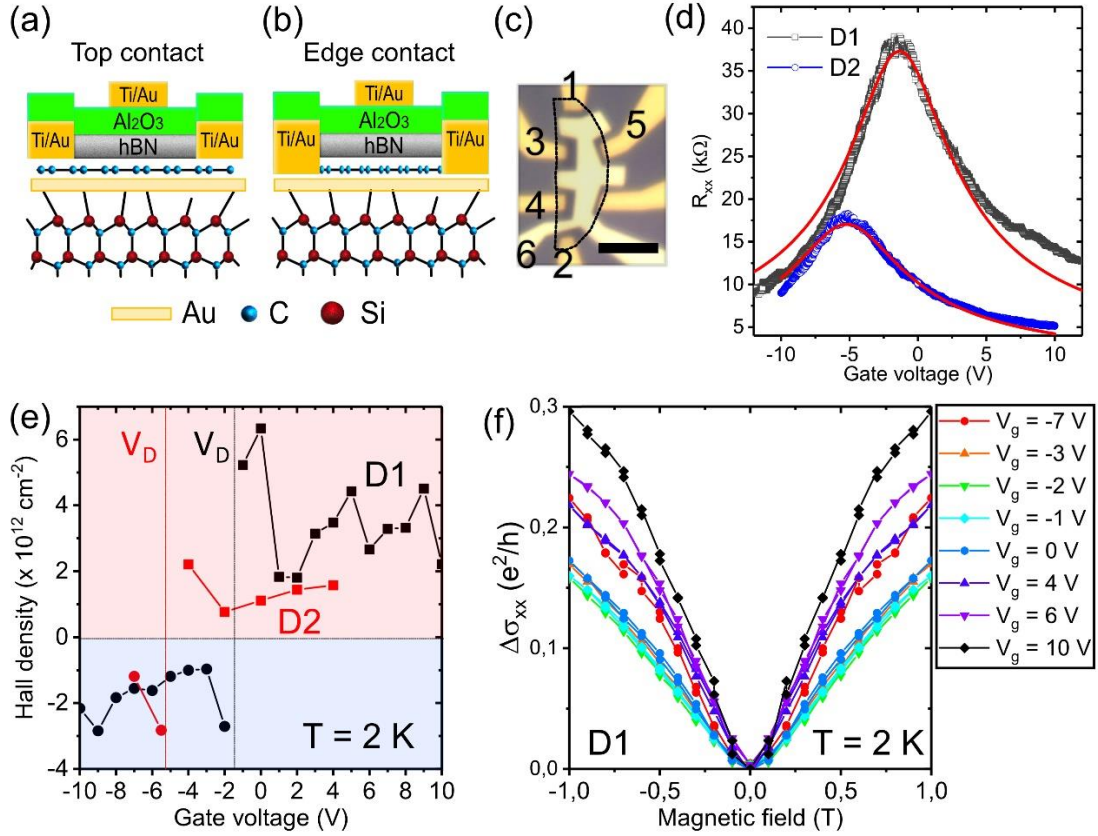

**Figure 3. Electron transport studies on Au-intercalated quasi-free-standing monolayer graphene.** (a,b) Schematics of top gate structures with h-BN/$Al_2O_3$ gate dielectrics: (a) top contact (D1) and (b) edge contact (D2) configurations. (c) An optical micrograph of D1. The dashed areas indicate the sample covered by h-BN. Measurement configuration: Current bias (1-2), $V_{xx}$ (3-4), $V_{xy}$ (5-3), gate (6-2). The scale bar is 5 μm. (d) Top gate voltage dependence of the two devices D1 and D2 at T = 2 K (I = 100 nA (D1) and 1 μA (D2)) showing Dirac points in both devices. The red lines are the fitted curves to the equation, $R_{xx} = \frac{1}{e\mu_C}\left(\frac{L}{W}\right)\frac{1}{\sqrt{n_g^2+n_0^2}}$. (e) Gate voltage dependence of Hall carrier density of D1 (black, I = 100 nA) and D2 (red, I = 1 μA) at T = 2 K, showing transition from electrons (red shaded area) to holes (blue shaded area) across the DP. (d) $\Delta\sigma_{xx}$ of D1 at T = 2 K (current bias I = 100 nA) at different gate voltages.

**Table 1.** Summary of Measured Devices

| Dev. | $t_{Au}$ (Å) | Region | Doping type | Carrier density ($10^{12}$ cm$^{-2}$) | Mobility (cm$^2$/Vs) | $V_D$ (V) |
|---|---|---|---|---|---|---|
| D1 | 4 | Au-diffused | Electron | 0.7 | 210 | -1.4 |
| D2 | 4 | Au-diffused | Electron | 1.2 | 560 | -5.2 |
| D3 | 4 | Au-deposited | Hole | 10 | 90 | - |
| D4 | 4 | Au-deposited | Hole | 10 | 80 | - |
| D5 | 8 | Au-diffused | Hole | 7 | 30 | - |
| D6 | 8 | Au-diffused | Hole | 5 | 90 | - |
| D7 | 8 | Au-deposited | Hole | 10 | 50 | - |
| D8 | 8 | Au-deposited | Hole | 20 | 15 | - |

**Supporting Information**

Material characterizations and transport data

**Corresponding Author**

*E-mail: samuel.lara@chalmers.se.

**Present Addresses**

†If an author's address is different than the one given in the affiliation line, this information may be included here.

**Author Contributions**

S.L.A. and S.K. conceived the experiment. R.Y. synthesized buffer layers. K.H.K., H.H., and S.L.A. carried out fabrication of devices, electron transport measurements, and analysis of transport data. C.S. and A.Z. performed synchrotron experiments and analyzed data. C.G. and A.T. carried out STM measurements and analyzed data. K.H.K. and S.L.A. drafted the manuscript and the manuscript was written through contributions of all authors. All authors have given approval to the final version of the manuscript.

**Funding Sources**

Swedish Foundation for Strategic Research (SSF) (No. IS14-0053, GMT14-0077, and RMA15-0024), Knut and Alice Wallenberg Foundation, Chalmers Area of Advance NANO, and the European Union's Horizon 2020 research and innovation programme Graphene Flagship under grant agreement No 785219.

**Acknowledgments**

This work was jointly supported by the Swedish Foundation for Strategic Research (SSF) (No. IS14-0053, GMT14-0077, and RMA15-0024), Knut and Alice Wallenberg Foundation, Chalmers Area of Advance NANO. This work was performed in part at Myfab Chalmers. This project has received funding from the European Union's Horizon 2020 research and innovation programme Graphene Flagship under grant agreement No 785219.

ABBREVIATIONS

Bu-L, buffer layer; MLG, monolayer graphene; SiC, Silicon carbide; SOI, spin orbit interaction; RMS, root mean square; FWHM, full width at half maximum; STM, scanning tunneling microscopy; LEEM, low energy electron microscopy; LEED, low energy electron diffraction; micro-XPS, x-ray photoelectron microspectroscopy; EBL, electron beam lithography; h-BN, hexagonal boron nitride; DP, Dirac point.

# Supplementary information for:

# Ambipolar charge transport in quasi-free-standing monolayer graphene on SiC obtained by gold intercalation

*Kyung Ho Kim,[1] Hans He,[1] Claudia Struzzi,[2] Alexei Zakharov,[2] Cristina Giusca,[3] Alexander Tzalenchuk,[3,4] Rositsa Yakimova,[5] Sergey Kubatkin,[1] Samuel Lara-Avila[1,3,*]*

[1]Department of Microtechnology and Nanoscience, Chalmers University of Technology, SE-412 96, Gothenburg, Sweden

[2]MAX IV Laboratory, 221 00, Lund, Sweden

[3]National Physical Laboratory, Hampton Road, Teddington TW11 0LW, UK

[4]Royal Holloway, University of London, Egham TW20 0EX, UK.

[5]Department of Physics, Chemistry and Biology, Linkoping University, 581 83 Linköping, Sweden.

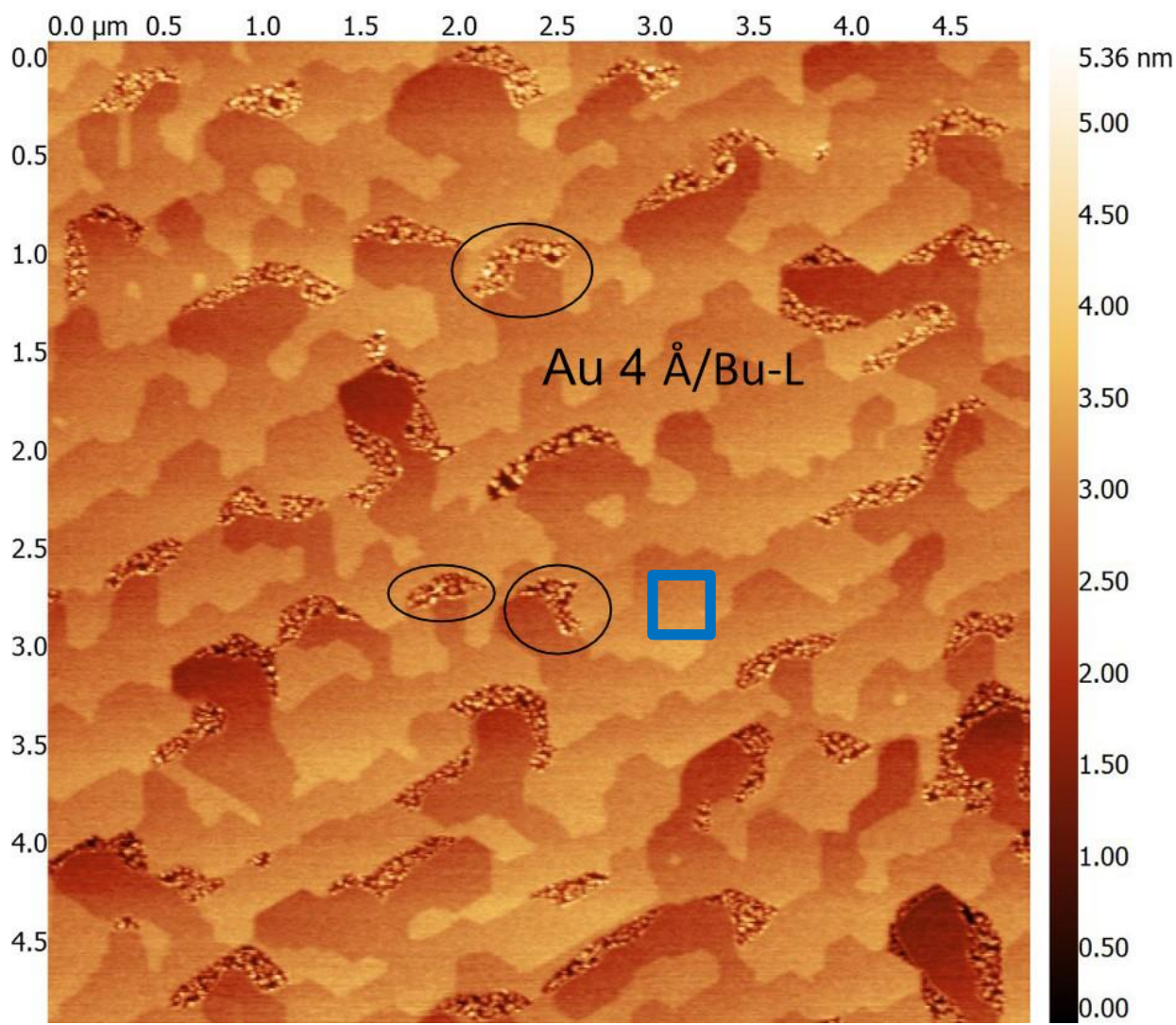

**Figure S1. Atomic force microscopy (AFM) of substrates after Au deposition.** (a) AFM topography of the Bu-L after 4 Å Au deposition (i.e. before Au intercalation at high temperature). The surface is smooth after deposition of Au (RMS roughness ~1 Å) in areas such as the one indicated by the blue rectangle. Au agglomeration only take place on the substrate areas where monolayer patches are present (circled areas). The height of SiC substrate steps is ~5 Å.

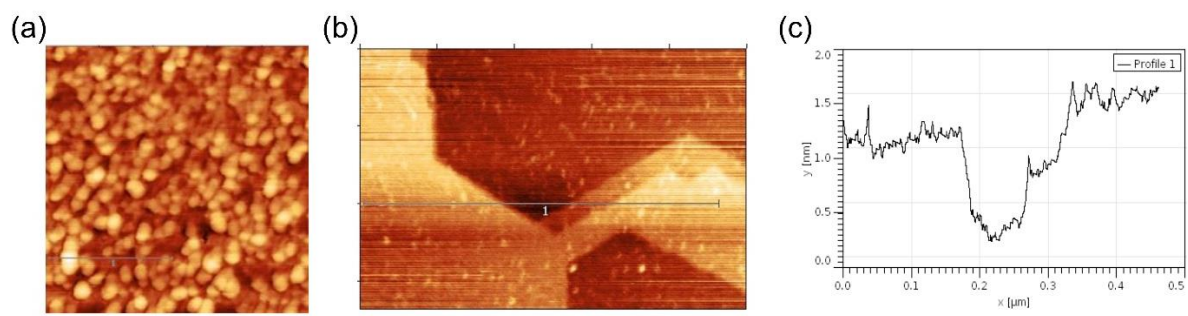

**Figure S2. Atomic force microscopy (AFM) of Au-intercalated quasi-free-standing monolayer graphene on SiC ($t_{Au}$ = 8 Å).** Atomic force microscopy image of the (a) Au-deposited area, and (b) the Au-diffused areas after thermal annealing at 850 °C. (c) Height profile in (b) showing atomically flat surface of intercalated surface.

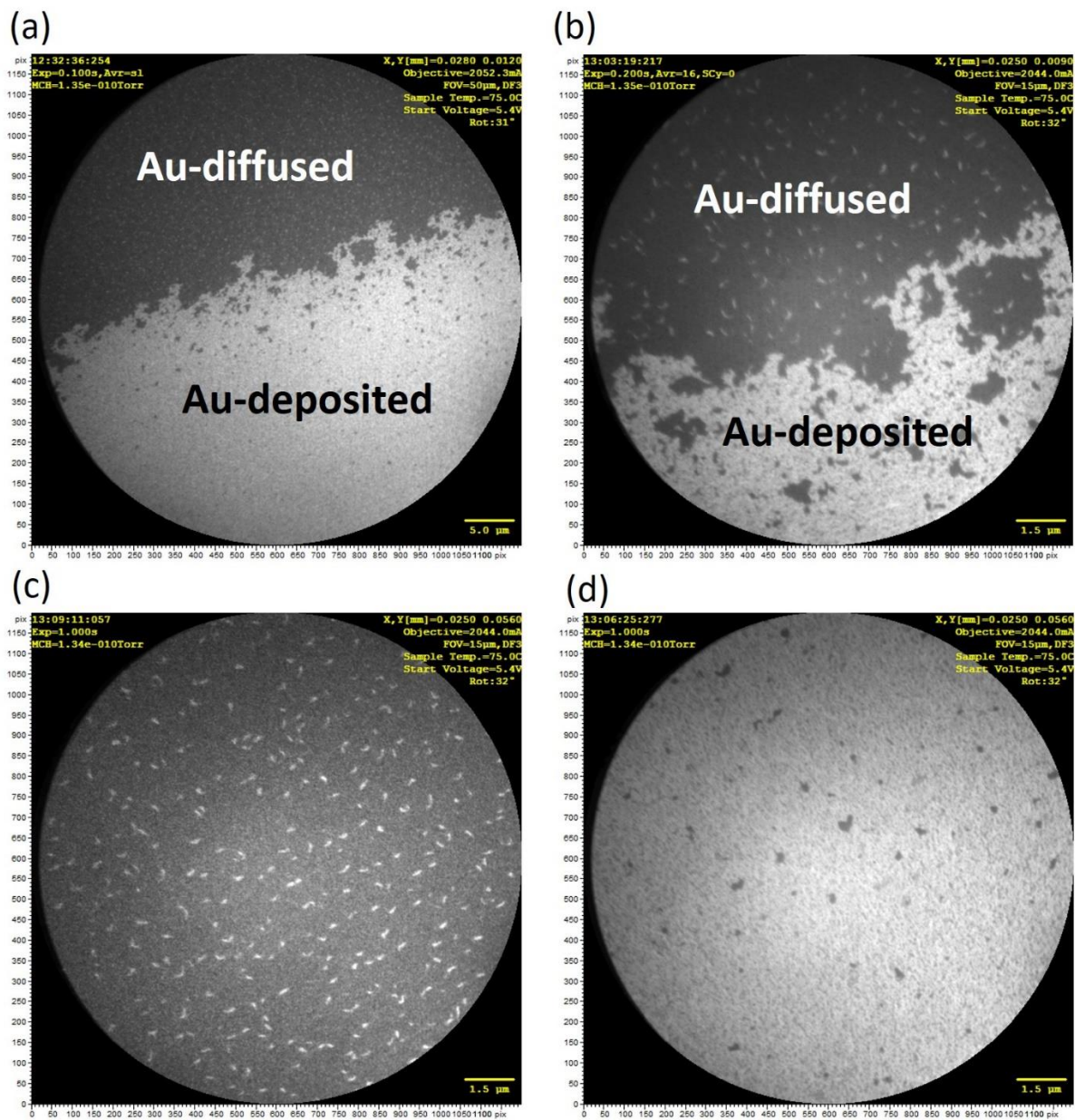

**Figure S3. Low energy electron microscopy (LEEM) of Au-intercalated quasi-free-standing monolayer graphene on SiC ($t_{Au}$ = 8 Å).** (a) LEEM image taken at the boundary of the deposited Au-deposited and Au-diffused areas. (b) The zoom in of (a) shows homogeneous morphology of quasi-free-standing monolayer graphene and bilayer graphene patches in the Au-diffused (c) and Au-deposited (d) areas.

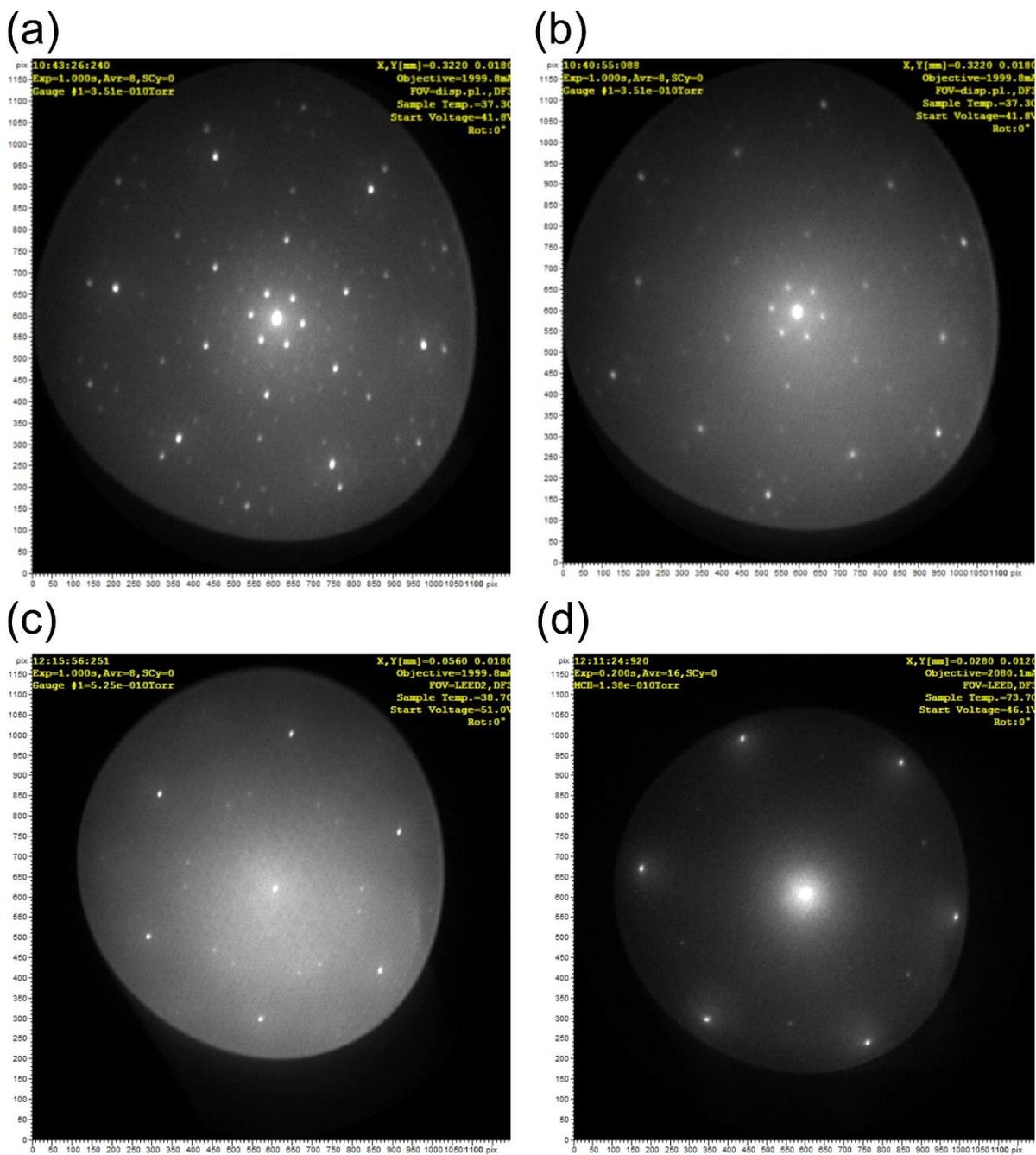

**Figure S4**. **Low energy electron diffraction (LEED).** (a) diffractogram of pristine Bu-L and (b) the Au-deposited area ($t_{Au}$ = 8 Å) before Au intercalation show the $6\sqrt{3} \times 6\sqrt{3}$ R30 grid, which is the characteristic of Bu-L. After thermal intercalation, the $6\sqrt{3} \times 6\sqrt{3}$ R30º grid disappear

and the crystalline structure of monolayer graphene appears for both (c) Au-deposited and (d) Au-diffused regions,

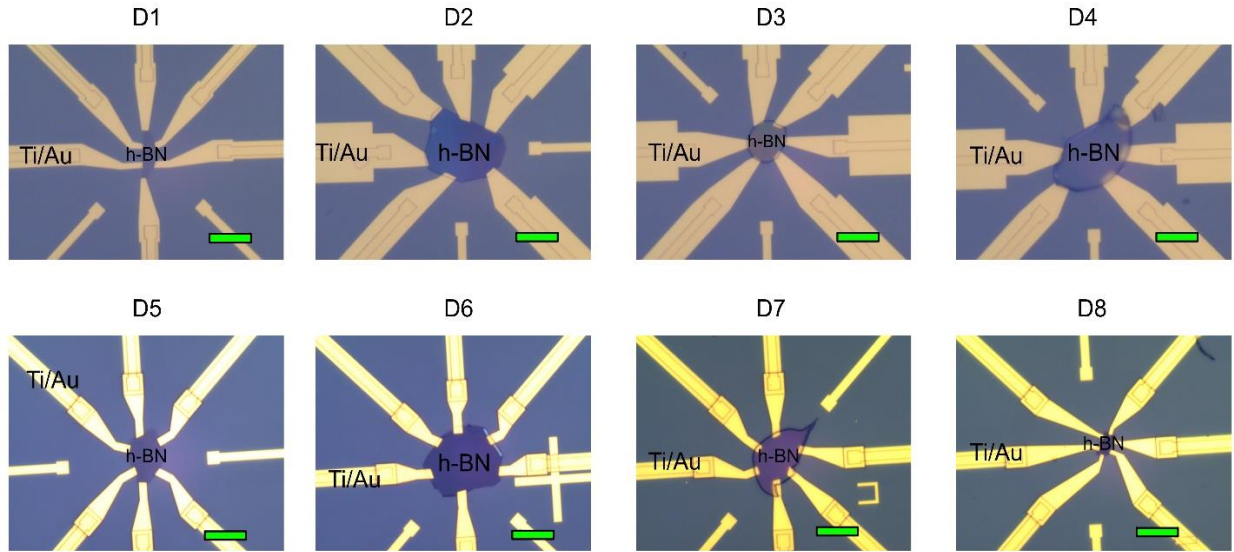

**Figure S5 Collection of optical photographs of devices D1 to D8 before fabricating top gate.** The shape of the Au intercalated Bu-L is dictated by the shape of h-BN and Ti/Au contacts are extended to samples. D2 is the edge contact and the others are top contact devices (See Fig. 3a,b). The scale bar is 10 µm.

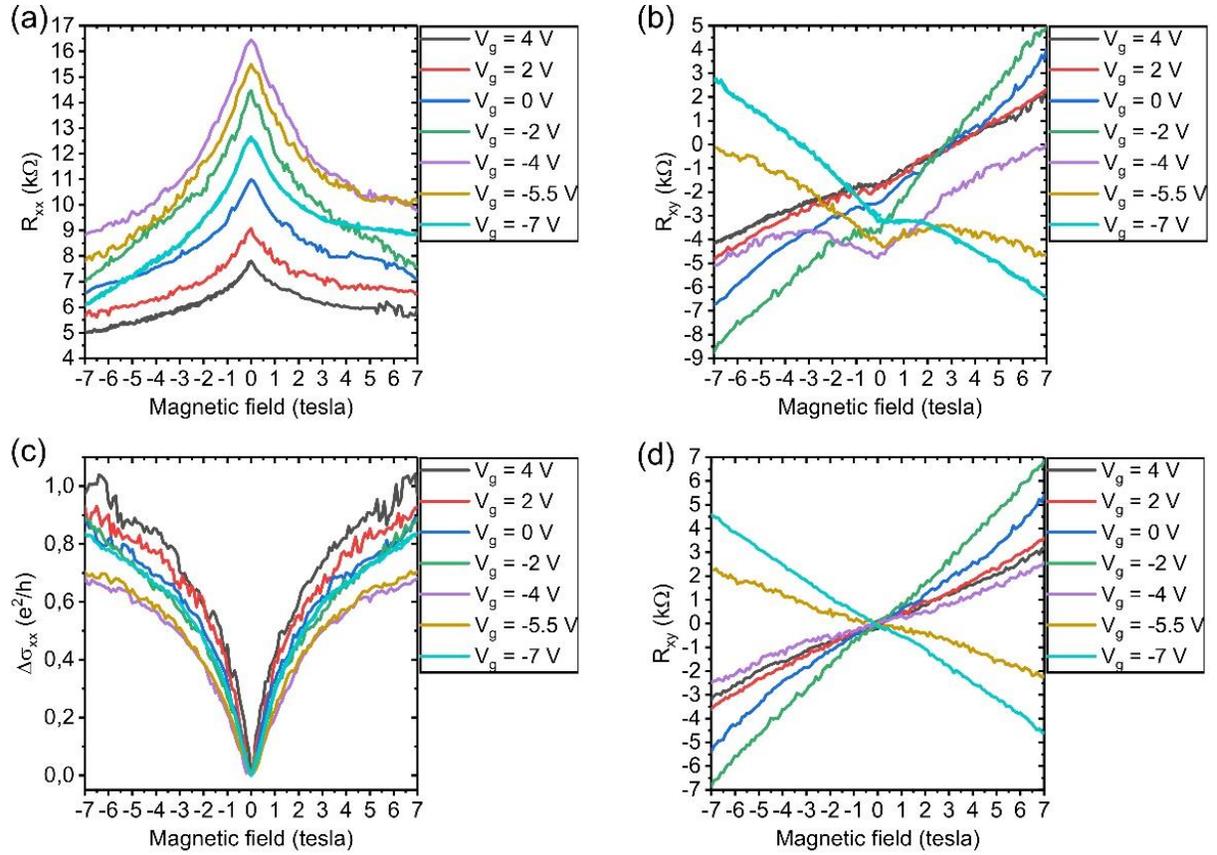

**Figure S6 Magnetotransport in Au intercalated quasi-free-standing monolayer graphene.**
(a) Magnetic field dependence of (a) longitudinal resistance $R_{xx}$ and (b) transverse resistance $R_{xy}$ of D2 at different top gate voltages $V_g$. (c) Symmetrized $\Delta\sigma_{xx} = \sigma_{xx}(B) - \sigma_{xx}(0)$, where we calculated $\sigma_{xx} = 1/R_{xx}$ and $\sigma_{xx}$ is symmetrized from data (a). (d) Asymmetric parts of $R_{xy}$ extracted from data (b).

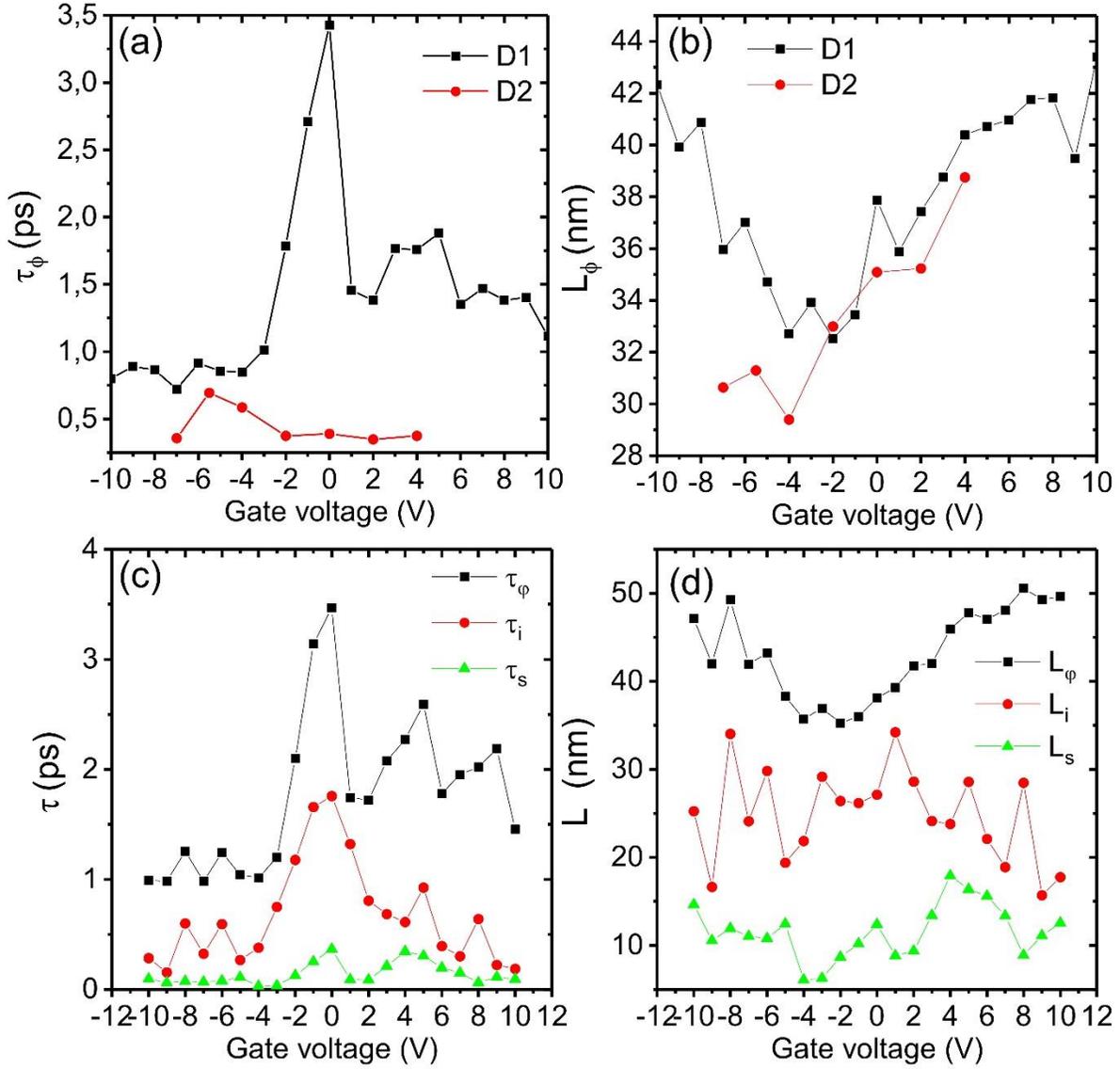

**Figure S7. Weak localization analysis for Au intercalated quasi-free-standing monolayer graphene.** (a) Phase coherence time in D1 and D2 obtained from the curvature of magneto conductivity at B = 0 and (b) corresponding phase relaxation lengths ($L_\varphi$) of D1 and D2. Scattering times $\tau_\varphi$, $\tau_i$, and $\tau_*$ (c) and scattering lengths $L_\varphi$, $L_i$, and $L_s$ of D1 (d) extracted from the best fits to the WL equation of monolayer graphene,[1]

$$\Delta\sigma_{xx} = \frac{e^2}{\pi h}\left[F\left(\frac{B}{B_\varphi}\right) - F\left(\frac{B}{B_\varphi + 2B_i}\right) - 2F(\frac{B}{B_\varphi + B_*})\right]$$

, where $F(x) = \ln(x) + \psi\left(\frac{1}{2} + \frac{1}{x}\right)$, $B_{\varphi,i,*} = \frac{\hbar c}{4De\tau_{\varphi,i,*}}$, $\ln(x)$ the natural logarithmic function, $\psi(x)$ the digamma function, D the diffusion coefficient, and $\tau_{\varphi,i,*}^{-1}$ the phase relaxation rate, inter-valley scattering rate, and inter/intravalley scattering rate including trigonal warping and chirality breaking, respectively, and $L_{\varphi,I,*} = (D\tau_{\varphi,i,*})^{1/2}$

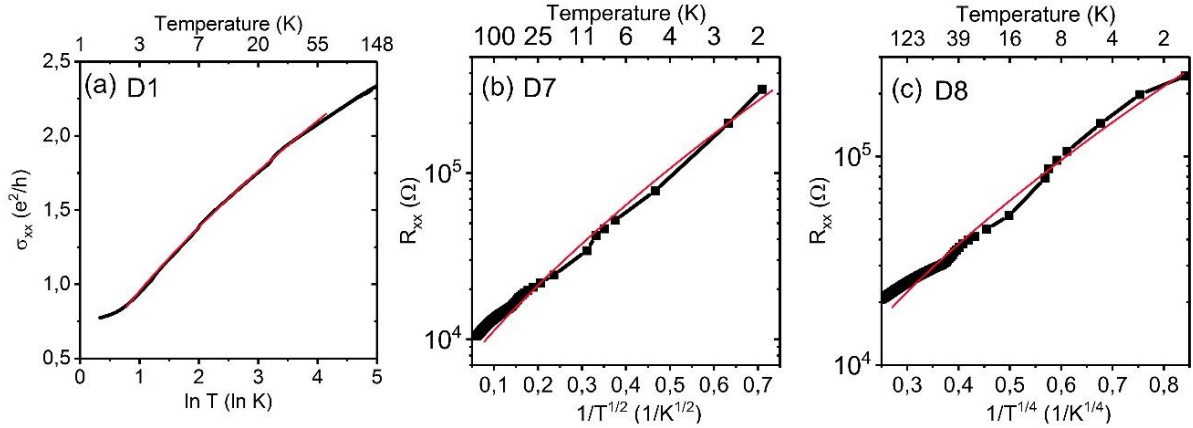

**Figure S8 Temperature dependence of conductivity and resistance for Au intercalated quasi-free-standing monolayer graphene.** (a) D1 fabricated on the Au-diffused area ($t_{Au}$ = 4 Å) close to ln(T), (b) D7 fabricated on Au-deposited area ($t_{Au}$ = 8 Å), $R(T) = R_0 \exp(\frac{T_0}{T})^{1/2}$, (c) D8 fabricated on Au-deposited area ($t_{Au}$ = 8 Å), $R(T) = R_0 \exp(\frac{T_0}{T})^{\frac{1}{4}}$ dependence as a function of temperature T. Red linear lines are guides to eyes.